\documentclass[times, review, 10pt]{elsarticle}

\usepackage{amssymb}
\usepackage{amsmath}

\usepackage{lineno} 

\usepackage{url}
\usepackage{xcolor}
\usepackage{algorithm}
\usepackage{algpseudocode}
\usepackage{booktabs}
\usepackage{multirow}
\usepackage{graphicx}
\usepackage{float}
\usepackage[normalem]{ulem}
\useunder{\uline}{\ul}{}
\journal{}

\begin{document}

\begin{frontmatter}



\title{Improving Lesion Segmentation in Medical Images by Global and Regional Feature Compensation} 


\author[label1]{Chuhan Wang}
\ead{chuhan.wang@sydney.edu.au}
\author[label2]{Zhenghao Chen\corref{cor1}}
\ead{zhenghao.chen@newcastle.edu.au}
\author[label3]{Jean Y. H. Yang}
\ead{jean.yang@sydney.edu.au}
\author[label1]{Jinman Kim\corref{cor1}}
\ead{jinman.kim@sydney.edu.au}
\cortext[cor1]{Corresponding authors: 
  Email: jinman.kim@sydney.edu.au; zhenghao.chen@newcastle.edu.au}

\affiliation[label1]{organization={Biomedical Data Analysis and Visualisation (BDAV) Lab, School of Computer Science, Faculty of Engineering, The University of Sydney},
            city={Sydney},
            postcode={2008}, 
            state={NSW},
            country={Australia}}

\affiliation[label2]{organization={School of Information and Physical Sciences, The University of Newcastle},
            city={Newcastle},
            postcode={2308}, 
            state={NSW},
            country={Australia}}

\affiliation[label3]{organization={School of Mathematics and Statistics, Faculty of Science, The University of Sydney},
            city={Sydney},
            postcode={2006}, 
            state={NSW},
            country={Australia}}
\begin{abstract}
Automated lesion segmentation of medical images has made tremendous improvements in recent years due to deep learning advancements. However, accurately capturing fine-grained global and regional feature representations remains a challenge. Many existing methods achieve suboptimal performance in complex lesion segmentation due to information loss during typical downsampling operations and insufficient capture of either regional or global features. To address these issues, we propose the Global and Regional Compensation Segmentation Framework (GRCSF), which introduces two key innovations: the Global Compensation Unit (GCU) and the Region Compensation Unit (RCU). The proposed GCU addresses resolution loss in the U-shaped backbone by preserving global contextual features and fine-grained details during multiscale downsampling. Meanwhile, the RCU introduces a self-supervised learning (SSL) residual map generated by Masked Autoencoders (MAE), obtained as pixel-wise differences between reconstructed and original images, to highlight regions with potential lesions. These SSL residual maps guide precise lesion localization and segmentation through a patch-based cross-attention mechanism that integrates regional spatial and pixel-level features. Additionally, the RCU incorporates patch-level importance scoring to enhance feature fusion by leveraging global spatial information from the backbone. Experiments on three publicly available medical image segmentation datasets, including brain stroke lesion, lung tumor and coronary artery calcification datasets, demonstrate that our GRCSF outperforms state-of-the-art methods, confirming its effectiveness across diverse lesion types and its potential as a generalizable lesion segmentation solution. 
\end{abstract}




\end{frontmatter}

\section{Introduction}

In medical imaging, accurate lesion segmentation plays a critical role in assisting clinicians with diagnosing diseases, monitoring disease progression, and evaluating treatment effectiveness~\cite{bib1}. Despite advancements in automated segmentation methods, accurately localizing and delineating lesions with complex characteristics remains challenging. For example, ischemic stroke lesions in the brain pose a significant segmentation challenge due to their low-contrast, where the pixel intensities are similar to those of the surrounding tissues, referred to as isointense~\cite{bib51} in MR imaging. The limitations of T1-weighted imaging to capture edema and early ischemic changes~\cite{bib3} make low-contrast stroke lesions difficult to distinguish from normal tissues. Similar to brain lesions, non-small-cell lung cancer (NSCLC) tumors on thin-section chest CT, despite their generally higher contrast, remain difficult to delineate accurately because their shapes, sizes, and locations vary markedly across patients. These lesions often exhibit irregular and ambiguous boundaries, further complicating their segmentation with traditional deep learning methods. Another example of challenging segmentation task is coronary artery calcification (CACS), occupying only a small region of the image, often measuring less than 5 mm in length and corresponding to 10–13 pixels referring to different in-plane resolution in the non-contrast CT slices used in this study~\cite{bib5}. The small size and irregular shape of the CACS, along with its variability in location and density, make precise segmentation and quantification difficult. Furthermore, non-contrast CT, the primary imaging modality for detecting and assessing CACS, further complicates the process due to its low signal-to-noise ratio, resulting in blurred boundaries and imprecise delineation.

Automated segmentation methods have improved significantly in recent years, driven by advances in deep learning~\cite{bib43}. Convolutional neural networks (CNN), particularly encoder-decoder architectures such as U-Net~\cite{bib12} and UNet ++~\cite{bib13}, have been widely applied for medical image segmentation task. These architectures preserve regional context but failed to capture global context due to fixed-size convolutional kernels, limiting their ability to segment lesions with varying locations or ambiguous boundaries. To address these limitations, advanced CNN-based models~\cite{bib66, bib67, bib68}, such as DeepLabv3~\cite{bib14} implement various receptive fields to improve the segmentation of complex anatomical structures, but remain limited by the fundamental constraints of convolutional operations, particularly in modeling long-range dependencies. 

In recent years, transformer-based models have demonstrated their potential in addressing these limitations by modeling global context and capturing long-range dependencies through self-attention mechanisms. TransUNet~\cite{bib17}, TransFuse~\cite{bib65},  UTNet~\cite{bib64} and SwinUNet~\cite{bib18} integrate hybrid architectures to learn regional and global characteristics, resulting in improved segmentation performance. However, they rely on skip connections to transfer low-level features from the encoder to the decoder for spatial detail recovery. This approach may not fully capture fine-grained details, limiting their localization ability and accuracy in segmenting challenging lesions. Several other transformer-based models~\cite{bib48} have been developed to solve the challenges of medical image segmentation. While these models have shown notable improvements in handling complex and variable structures, they still face challenges in accurately localizing subtle and irregular lesions and delineating their boundaries. This limitation arises because patch-based self-attention tends to blur regional details and introduce feature inconsistencies among patches. SPiN~\cite{bib19} is developed to address these challenges using the subpixel mechanism. Although effective in handling small stroke lesion segmentation, SPiN relies heavily on supervised learning approach and requires extensive manually annotated datasets.

Recent methods have enhanced performance in segmentation tasks by employing self-supervised learning (SSL), which learns meaningful representations from unlabeled data. SSL reduces reliance on costly and time-intensive manual annotations while improving downstream task performance. It achieves this through pretext tasks that exploit the inherent structure of images, such as contrastive learning~\cite{bib21}, pixel-level image reconstruction~\cite{bib24, bib54}, and masked token prediction~\cite{bib25}. These tasks enable models to extract robust global features~\cite{bib47} from datasets and localized features from individual images, which can serve as initialization weights for supervised learning. SSL has demonstrated particular effectiveness in medical imaging~\cite{bib28}, showing that SSL pre-training improves segmentation performance, especially when fine-tuned with small annotated datasets. However, many SSL methods focus predominantly on learning high-level global features that capture the overall data context, often overlooking fine-grained, pixel-level details. These details, critical for segmentation tasks, such as precisely delineating the boundaries of challenging lesions with low-contrast or small sizes, remain unexplored in existing SSL approaches. This gap underscores the need for SSL methods that integrate global and pixel-level information to enhance segmentation performance. Masked Autoencoders (MAE)~\cite{bib24}, introduced by He {\it \textit{et al.,}} addresses this limitation by incorporating localized feature representations. Although MAE has been used primarily as a pre-training technique in medical imaging~\cite{bib30, bib31}, its potential for tasks beyond pre-training, such as using reconstructed images to guide segmentation remains unexplored.

Foundation models such as the Segment Anything Model (SAM)~\cite{bib55} and GPT-4 with vision (GPT-4V)~\cite{bib56} have emerged as powerful tools for enabling zero-shot and few-shot segmentation through user-provided prompts. SAM performs segmentation based on prompts such as points, bounding boxes, or masks. Trained on over 1 billion masks across 11 million images, SAM can deliver competitive or even superior performance compared to traditional supervised models in a zero-shot or few-shot scenarios. However, its performance is highly dependent on the quality of the input prompts. Since SAM lacks the ability to correct prompts that entirely fall within false-positive regions, inaccurate or suboptimal prompts can lead to poor segmentation results. GPT-4V, a vision-enhanced large language model (LLM), incorporates image inputs and opens new perspectives for generating image-text pairs. Despite its potential, as it is not specifically trained for the medical domain or for segmentation tasks, GPT-4V may understand general content in medical images but lacks the fine-grained or pixel-level understanding needed to localize complex anatomical structures and lesions~\cite{bib57}. As a result, it cannot reliably generate accurate bounding boxes or masks as prompts for downstream segmentation models, nor can it serve directly as a segmentation model due to the absence of a dedicated segmentation head.

To address the limitations mentioned above, we propose a novel dual-feature compensation framework, named Global and Regional Compensation Segmentation Framework (GRCSF), to improve the medical image segmentation of challenging lesions. GRCSF adopts an encoder-decoder structure built on UNet++ and takes the raw images as input. It compensates for the detailed global features that are often missing in CNN-based models. Unlike traditional approaches that use MAE for pre-training, our method, to our knowledge, is the first medical image segmentation framework to use MAE’s reconstruction process to generate SSL residual maps that capture pixel-level differences between the original images and their reconstructions. Therefore, this mitigates inconsistencies between global and local feature representations introduced by transformer models. The key contributions of our work are as follows:

\begin{itemize}
    
\item We introduce a novel medical image segmentation framework, namely, GRCSF, which exploits global and regional feature compensation strategies to produce better coarse-to-fine features and improve lesion segmentation.

\item  We introduce the Global Compensation Unit (GCU) module within the encoder, using feature recovery from the similarity of different scales. It updates skip connection features by using the global context and progressively refines the multi-scale features. It addresses the loss of detailed global features caused by downsampling and solves the problems of inaccurate boundary delineation and low sensitivity in segmenting complex lesions.

\item  We introduce the Regional Compensation Unit (RCU) module within the decoder, by applying a cross-attention mechanism between residual maps learned from the SSL strategy and the features of UNet++. We can obtain importance score maps to better highlight those regions with a higher likelihood of containing lesions by incorporating additional global features from the backbone model, which helps reduce false positives in SSL residual maps. The RCU addresses the challenge of localizing blurred lesions, which are often difficult to detect due to limited regional spatial features.

\item  We validate the effectiveness and robustness of GRCSF for general lesion segmentation, using three representative datasets: ATLAS 2.0, including low-contrast brain stroke lesions; MSD Lung Tumor Segmentation, featuring irregular NSCLC tumors; and orCaScore, which contains small-sized coronary artery calcifications.  
\end{itemize}

\begin{figure*}[htbp]
    \centering
    \includegraphics[scale=0.35]{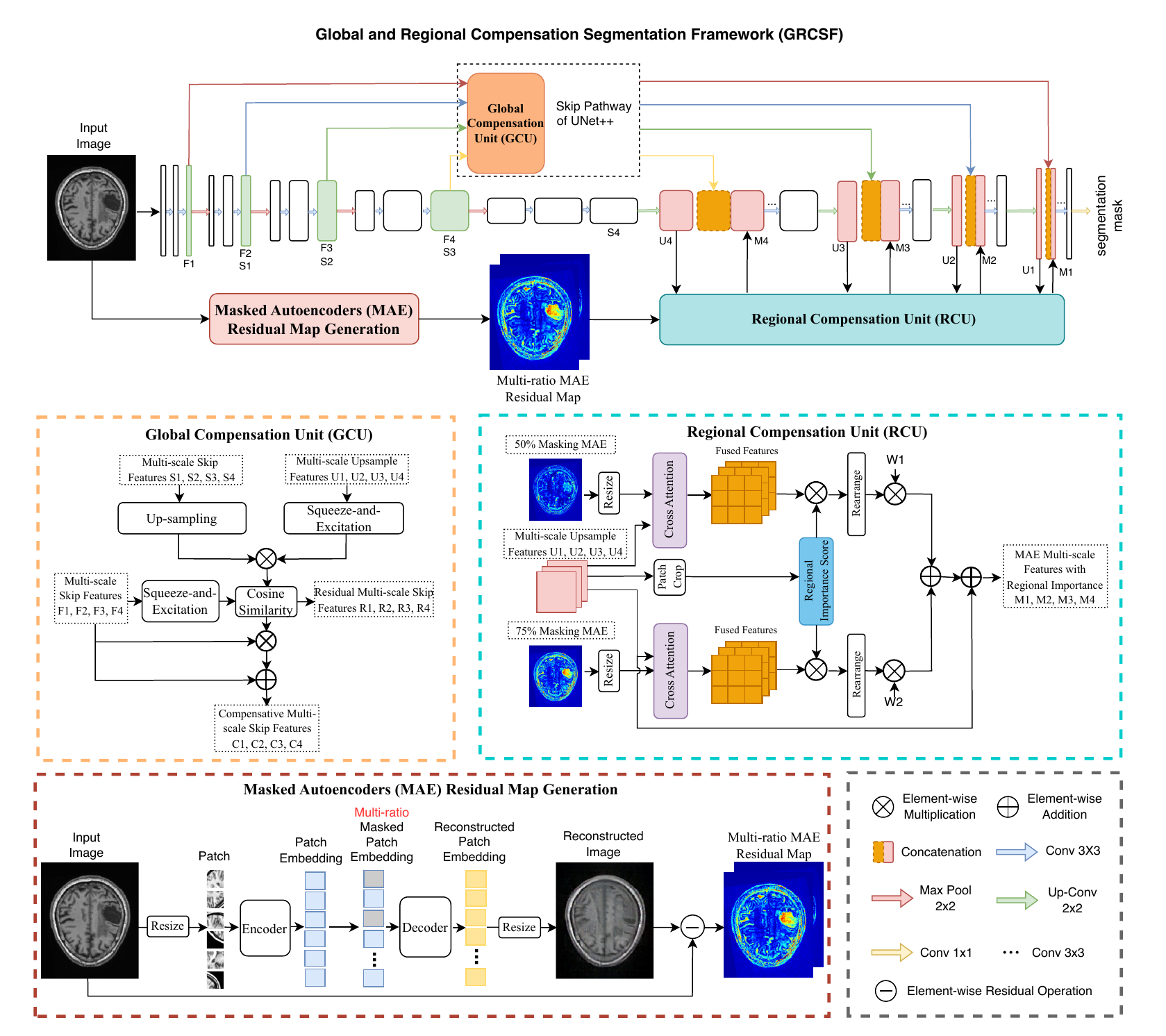}
    \caption{Overview of the GRCSF architecture. A U-shaped backbone network for medical image segmentation. The backbone integrates the Global Compensation Unit (GCU) within the skip connections and the Regional Compensation Unit (RCU) in the decoder layers. The masked Autoencoders (MAE) model was implemented using pre-trained weights for input size of 224x224. To ensure the correct input and output sizes, we resized the input images before and after processing through the MAE.}
    \label{Fig:1}
\end{figure*}

\section{Related Work}

\subsection{Deep Learning for Supervised Medical Image Segmentation}

The CNN-based encoder-decoder architectures such as U-Net~\cite{bib12} and its variant UNet ++~\cite{bib13}, have been widely adopted for the medical image segmentation task. These architectures utilize skip connections to preserve spatial resolution, making them effective for various segmentation tasks. Unet++ improved multi-scale feature extraction by introducing nested skip pathways, further enhancing its ability to handle complex and heterogeneous medical data. DeepLabv3~\cite{bib14} introduced Atrous Spatial Pyramid Pooling (ASPP)~\cite{bib15} to encode multi-scale contextual information by probing features at multiple rates and fields of view. A recent study SPiN~\cite{bib19} has made efforts to accurately segment isointense brain strok lesions. It developed a subpixel embedding mechanism to generate high-resolution confidence maps and employed a learnable downsampler to accurately localize irregular lesions and delineate boundaries. While CNN-based segmentation models have demonstrated improvement in medical image segmentation, they are still limited to capture global features due to fixed receptive field.

Transformer-based models TransUNet~\cite{bib17}, TransFuse~\cite{bib65} and UTNet~\cite{bib64} combined CNNs and transformers in a hybrid architecture, leveraging CNNs for high-resolution spatial encoding and transformers for capturing long-range dependencies. Similarly, SwinUNet~\cite{bib18} adopted a fully transformer-based U-shaped model, incorporating a hierarchical Swin Transformer~\cite{bib52} with a shifted windows mechanism as the encoder to simultaneously model global and local dependencies. However, the use of window-based attention in transformer-based models can introduce boundary discontinuities, which is particularly problematic for lesions with ambiguous edges. Moreover, its performance is heavily dependent on large training datasets and extensive parameter tuning, making adaptation to new datasets challenging. 

Existing methods either fail to effectively capture global features or lack of consistent regional feature representation. GRCSF utilizes pixel-level similarity from different feature scales to enhance feature representations.

\subsection{Self-Supervised Learning for Medical Image Segmentation}
SSL leverages unlabeled data to learn meaningful feature representations for different tasks. For example, MAE, originally developed for natural images, have also shown potential in medical imaging by reconstructing masked image patches and learning global semantic and localized contextual features. MAE employs an encoder-decoder architecture in which a portion of image patches are randomly masked during training. The encoder extracts semantic information from visible patches, while the decoder reconstructs the masked regions. This approach allows MAE to learn the overall structure of the dataset and the unique fine-grained details of each image. Zhou {\it \textit{et al.}}~\cite{bib30} has shown that MAE self pre-training improves various medical image tasks, including abdominal CT multi-organ segmentation, and MRI brain tumor segmentation. Mask in Mask (MiM)~\cite{bib31} framework introduced hierarchical token learning and cross-level alignment mechanisms to improve downstream segmentation tasks across large-scale datasets.

These advances highlight the impact of SSL in medical image segmentation, enabling models to leverage unlabeled data effectively and generalize across diverse medical imaging tasks. However, most SSL methods remain primarily focused on pre-training. To our best knowledge, GRCSF is the first medical image segmentation work to exploit the capacity of SSL to recover global features and produce residual maps that highlight the regions of potential lesion used for regional feature compensation. This operation takes advantage of additional features to improve the likelihood at the pixel-level of accurately identifying abnormal regions.

\subsection{Transformer-based Vision Foundation Models for Medical Image Segmentation}
Recent developments in large-scale vision foundation models have significantly influenced medical image segmentation. SAM ~\cite{bib55} is a representative prompt-driven segmentation framework and has demonstrated robust zero-shot performance across diverse medical imaging modalities ~\cite{bib58}. However, its effectiveness is highly dependent on the quality of the input prompts. To address this, previous studies ~\cite{bib60} have explored automated prompt generation strategies from coarse segmentation masks to enhance segmentation accuracy.

GPT-4V ~\cite{bib56} has shown its ability in modality recognition, disease diagnosis, medical visual question answering (VQA) and report generation. However, its performance is limited in certain tasks such as chest radiograph interpretation ~\cite{bib57} . Moreover, GPT-4V failed to localize the region-of-interest when it shared similar texture and shape with the background, highlighting its vulnerability to complex backgrounds ~\cite{bib61}. Although integrating such generalist models into clinical segmentation workflows remains a challenge, advances in prompt engineering ~\cite{bib62, bib63} offer a potential path toward improving localization accuracy and segmentation robustness.

\section{Method}
\subsection{Overview}
Our GRCSF is a dual-feature compensation framework based on an encoder-decoder architecture. The training process involves first using a pre-trained MAE model to process input images and generate two sets of SSL residual maps with mask ratios of 50\% and 75\%. Subsequently, the input images and their corresponding SSL residual maps are simultaneously processed through a UNet++ backbone to produce the final segmentation results. 

\begin{figure}[!t]
    \centerline{\includegraphics[width=0.5\columnwidth]{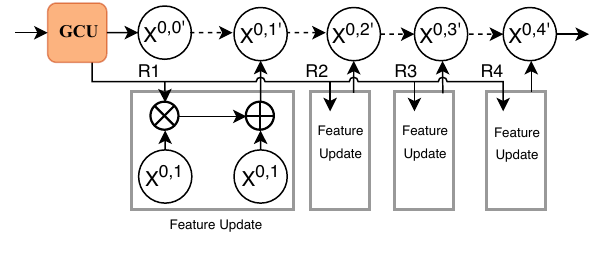}}
    \caption{ Details of skip pathway of UNet++. The skip connection features are updated using the GCU, where $X^{0,0’}$ represents the updated feature output from the GCU of the original skip feature $X^{0,0}$. Subsequently, $X^{0,1’}$, $X^{0,2’}$, $X^{0,3’}$, $X^{0,4’}$ are the updated features generated by incorporating the residuals from the GCU, followed by the operations illustrated in the figure. This process is applied to each skip concatenation layer in UNet++.}
    \label{Fig. 3}
\end{figure}

In the encoder, raw images are processed to extract hierarchical features. The GCU is introduced at each layer to recover pixel-wise information that is lost during downsampling. It does this by assigning higher weights to affected regions during feature concatenation between the encoder and the decoder, thereby enhancing the global feature representation. In the decoder, the RCU fuses multi-ratio residual maps (MRM) with upsampling features in the last three layers to highlight pixel-level discrepancies that likely correspond to lesions. The RCU processes these MRM using cross-attention and a patch-based mechanism that calculates importance scores, quantifying the likelihood that each patch contains lesions. By combining the complementary strengths of raw image features and MRM, GRCSF integrates global and regional information to improve segmentation accuracy, particularly for challenging lesions. The overall architecture is illustrated in Fig. \ref{Fig:1}.

\subsection{Global Compensation Unit}

The GCU addresses pixel-level information loss caused by downsampling in U-shaped CNN architectures by reintroducing lost details into the skip-concatenated features. During downsampling, the input feature map undergoes two convolutional operations, resulting in a downsampled feature map \(\mathit{S} \in \mathbb{R}^{H \times W \times C}\). To mitigate information loss, \(\mathit{S}\) is re-upsampled to match the resolution of the skip feature map from the previous layer, \(\mathit{F} \in \mathbb{R}^{H \times W \times C}\), producing \(\mathit{RU} \in \mathbb{R}^{H \times W \times C}\). A comparison between \(\mathit{RU}\) and \(\mathit{F}\) identifies regions of information loss.

To refine \(\mathit{RU}\), it is multiplied by the corresponding upsampled feature map \(\mathit{U} \in \mathbb{R}^{H \times W \times C}\) in the decoder, which is enhanced by a Squeeze-and-Excitation (\( \text{SE} \)) mechanism ~\cite{bib46}. The \(\text{SE}\) block, applied to both \(\mathit{F}\) and \(\mathit{U}\), focuses on critical regions while suppressing irrelevant areas. This process yields two key outputs:
The skip-concatenation residual map is defined as:
\[
R = \text{CS}(\mathit{RU} \otimes \text{SE}(U), \text{SE}(F)) \tag{1}
\]
The updated skip-concatenation features are defined as:
\[
C = \text{CS}(\mathit{RU} \otimes \text{SE}(U), \text{SE}(F)) \otimes \mathit{F} \oplus \mathit{F} \tag{2}
\]

Here, \(\text{SE}(\mathit{U})\) and \(\text{SE}(\mathit{F})\) are attention maps generated by the \(\text{SE}\) block. \text{CS} is the pixel-wise cosine similarity operation, while \( \otimes \) and \( \oplus\) denote the element-wise product and addition, respectively. Each position captures context along both horizontal and vertical axes. The residual map \(R\) highlights areas of significant change and is particularly relevant for architectures such as UNet++, where each layer involves multiple skip concatenation features. For simpler architectures such as UNet, the residual map \(R\) is not required, as the updated skip feature \(C\) alone is sufficient to perform skip concatenation between the encoder and decoder.

By improving the information flow (Fig. \ref{Fig. 3}) between the encoder and decoder, the GCU ensures better preservation of global and pixel-level details. It can be integrated into any U-shaped CNN, enhancing feature representations.

\subsection{Regional Compensation Unit}
The RCU is designed to generate a weighted feature map \( M \in \mathbb{R}^{H \times W \times C} \) by utilizing the MRM, combined with the upsampled feature \( U \). These MRM provide complementary information that enhances the ability of the segmentation network to focus on regions with a higher likelihood of containing lesions.

\subsubsection{Residual Maps from Self-Supervised Learning}
To compute the SSL residual maps of MRM, we adopt a SSL strategy with Mask Autoencoder illustrated in Fig. \ref{Fig:1}. We trained the input images with two different mask ratios: 50\% and 75\%. The 75\% mask ratio is used as the default, while the 50\% ratio is empirically chosen to capture sufficient image context. For each input image \(I \in \mathbb{R}^{H \times W \times 1}\), the MAE produces reconstructed images under both masking conditions. To reduce randomness caused by mask configurations that may fail to cover lesion areas, the MAE model is applied five times per mask ratio, generating five reconstructed images per input. The reconstructions, denoted as \(R_1^{(i)}\) and \(R_2^{(i)} \in \mathbb{R}^{H \times W \times 1} \, (i=1,\ldots,5)\), are averaged to produce more stable representations as \( R_1 = \frac{1}{5} \sum_{i=1}^{5} R_1^{(i)} \) and \( R_2 = \frac{1}{5} \sum_{i=1}^{5} R_2^{(i)} \).

\begin{figure*}[htbp]
    \centerline{\includegraphics[width=\textwidth]{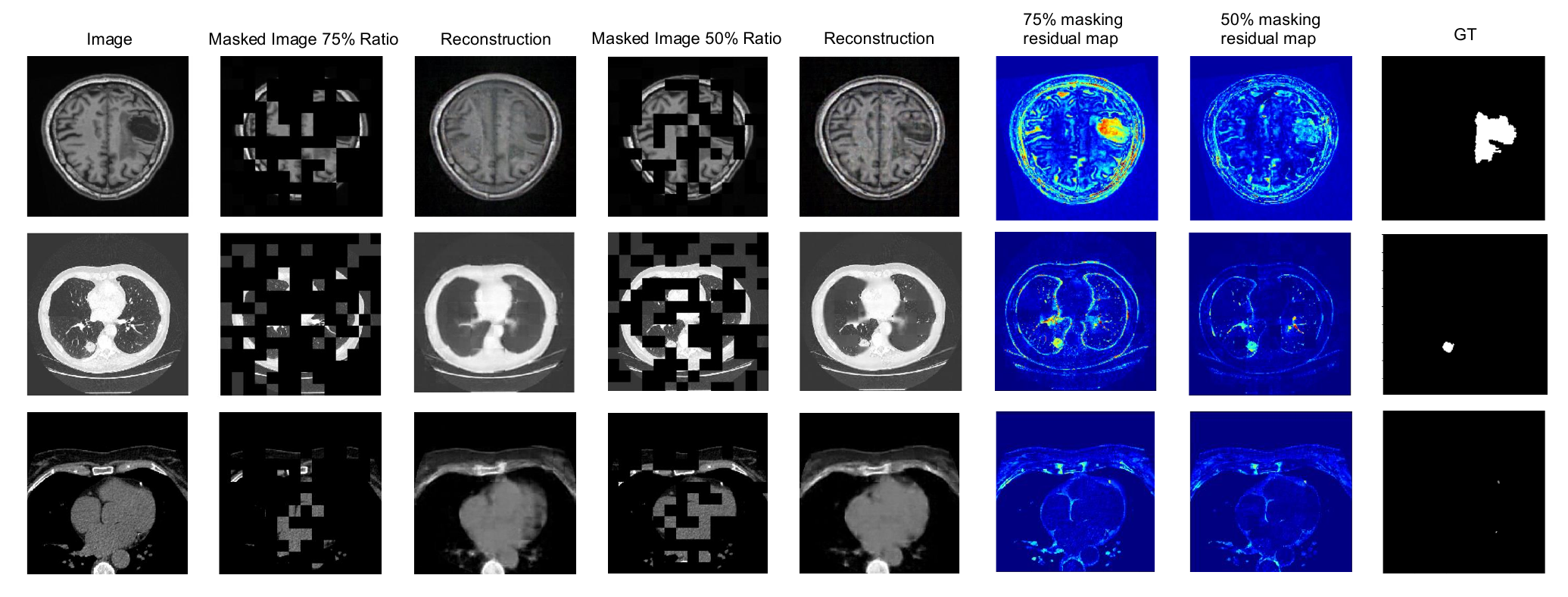}}
    \caption{Example visualizations of MAE reconstruction results for mask ratios of 75\% and 50\%. The figure includes the masked input images, reconstructed images, derived MRM highlighting the differences between the original images and their reconstructions, and the corresponding ground truth for reference. Please zoom in for a clearer view.}
    \label{Fig. 2}
\end{figure*}

\begin{figure}[!t]
    \centerline{\includegraphics[width=0.5\columnwidth]{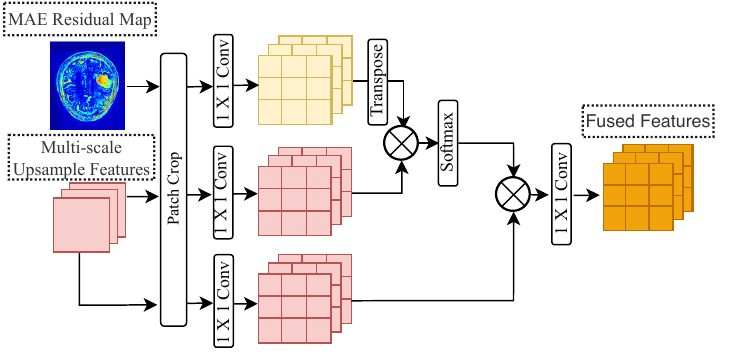}}
    \caption{Patch-based cross-attention mechanism in the RCU module. This mechanism aligns upsampling features from the decoder layer with MRM at the same feature scales. The outputs are fused feature maps that combine information from both inputs.}
    \label{Fig. 4}
\end{figure}

\begin{figure}[!t]
\centerline{\includegraphics[scale=0.7]{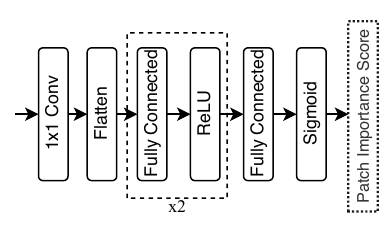}}
\caption{Regional importance scoring mechanism in the RCU module. This mechanism calculates the patch-based likelihood of lesion presence by processing decoder features generated by the backbone at each scale.}
\label{Fig. 5}
\end{figure}

The SSL residual maps \(RM_1\) and \(RM_2\) are computed by calculating the pixel-wise absolute difference (AbsDiff) between the input image \(\mathit{I}\) and the averaged reconstructions \(R_1\) and \(R_2\), respectively. The pixel-wise AbsDiff between two images is defined as:
\[
\text{AbsDiff}(I, R) = \left| \mathit{I}(x,y) - \mathit{R}(x,y) \right| \tag{3}
\] 
\[
\mathit{RM}_1 = \text{AbsDiff}(I, R_1) \tag{4}
\] 
\[
\mathit{RM}_2 = \text{AbsDiff}(I, R_2) \tag{5}
\]
where (x, y) are the pixel coordinates. This metric quantifies the AbsDiff between two arrays at corresponding pixel locations, representing their similarity on a scale from 0 to 1, where 0 indicates perfect alignment and 1 indicates complete opposition. These MRM highlight regions where the reconstructions differ from the original image, reflecting potential lesion areas (Fig. \ref{Fig. 2}).

\subsubsection{Regional Feature Fusion}
To compute the weighted feature \( M \), the RCU operates in a patch-based manner. The SSL residual maps \( RM_1 \) and \( RM_2 \) are divided into patches, resulting in patch-based representations \noindent ${RM}_1^{\prime} \in \mathbb{R}^{N \times P^2 \times C} \quad \text{and} \quad {RM}_2^{\prime} \in \mathbb{R}^{N \times P^2 \times C}$, where \( N = \frac{H}{P} \times \frac{W}{P} \) is the number of patches and \( P \times P \) is the size of each patch. Similarly, the upsampled feature map \( U \) is reshaped into a patch-based representation \( U^{\prime} \in \mathbb{R}^{N \times P^2 \times C} \). Patch-based cross-attention is applied between \noindent ${RM}_1^{\prime}$, ${RM}_2^{\prime}$, and ${U}^{\prime}$, respectively, enabling the MRM to highlight spatially relevant areas within ${U}^{\prime}$. Cross-attention operations are represented as \( ({RM}_1^{\prime} \circledast {U}^{\prime}) \) and \( ({RM}_2^{\prime} \circledast {U}^{\prime}) \), where \( \circledast \) denotes the cross-attention mechanism described in ~\cite{bib35} illustrated in Fig. \ref{Fig. 4}.

Simultaneously, the patches in ${U}^{\prime}$ are processed through an importance score module shown in Fig. \ref{Fig. 5}, which generates a score for each imaging feature patch as $\phi({U}^{\prime})$, where \( \phi \) consists of a \( 1 \times 1 \) convolution followed by two sets of fully connected layers with ReLU activations, an additional fully connected layer, and a sigmoid activation since multiple patches could contain lesions.

The cross-attention outputs are scaled by their corresponding importance scores and combined using learnable scalar weights \( W_1 \) and \( W_2 \) to produce the final feature map \( M \). The computation is given by:
\begin{align}
    M =  \varphi (({RM}_1^{\prime} \circledast {U}^{\prime}) \cdot \phi({U}^{\prime})) \cdot W_1 \oplus \notag \\ 
    \varphi (({RM}_2^{\prime} \circledast {U}^{\prime}) \cdot \phi({U}^{\prime})) \cdot W_2 \oplus U \tag{6}
\end{align}
where $\varphi$ is the operation to rearrange the feature map patches back to the original size in the backbone. 

The weighted feature map \( M \) replaces \( U \) in the backbone segmentation network. Importance scores guide the network’s attention to patches that are more likely to contain lesions, while the patch-based cross-attention mechanism further refines these patches by emphasizing lesion-specific regional features.

\subsection{Deployment and Objective Function}
The GCU and RCU modules can be integrated into any U-shaped CNN architecture that utilizes skip connections. Both modules offer flexibility and can be incorporated at various skip concatenation points or decoder layers, adapting to the specific needs of the backbone architecture and dataset characteristics. For the skip connections in the UNet++ backbone, the residuals from the first skip feature are added to subsequent skip connections (Fig. 2). The RCU generates a weighted feature map, which replaces the original upsampled feature map at the corresponding decoder layer.

The framework operates sequentially, with the MAE model first generating MRM from the input images. These MRMs are then used as additional information for the segmentation network, alongside the raw input images. The images are processed slice-wise, and the network outputs a single-channel probability map matching the input resolution after a sigmoid activation. The patch size of the MRM and the feature maps used in cross-attention are treated as tunable hyper-parameters. The MAE is trained using the mean squared error \textit{(MSE)} loss by default. For the segmentation network, a mixed loss function combining dice loss and focal loss is utilized for the ATLAS and the MSD Lung Tumor datasets, while focal loss is applied for the orCaScore dataset. This approach balances segmentation precision and recall, enhances sensitivity to challenging lesions, and addresses class imbalance. The mixed loss and focal loss are defined as follows:
\[
\text{Loss}_{\text{ATLAS}}^{\text{mix}} = 1 - \frac{2|x \cap y|}{|x| + |y|} - \alpha (1 - p_t)^\gamma \log(p_t) \tag{7}
\]
\[
\text{Loss}_{\text{orCaScore}}^{\text{focal}} = -\alpha (1 - p_t)^\gamma \log(p_t) \tag{8}
\]
where \( x \) and \( y \) are the target pixels in the predicted and ground truth (GT) images. \( \alpha \) is a weighting factor, \( p_t \in [0, 1] \) is the model’s estimated probability, \( \gamma \) is a focusing parameter, and we set \( \alpha = 0.25 \) and \( \gamma = 2 \) for both datasets.

This integration of the GCU and RCU into a U-shaped CNN backbone enhances the network’s ability to capture fine-grained details, improving sensitivity to subtle and small lesions. The flexible design can easily be adaptable to other encoder-decoder backbone architectures, making the approach widely applicable.

\section{Experiments}
\subsection{Datasets}
We used three publicly available medical image segmentation datasets as examples to evaluate lesions with low-contrast and small-size characteristics: the Anatomical Tracings of Lesions After Stroke (ATLAS) v2.0~\cite{bib32} dataset, the Medical Segmentation Decathlon (MSD) Lung Tumor Segmentation dataset~\cite{bib53} and the Coronary Artery Calcium Scoring (orCaScore)~\cite{bib33} dataset.

To evaluate segmentation performance on low-contrast lesions, we utilized the ATLAS 2.0 dataset, which comprises 955 samples of 655 public training cases and 300 hidden test cases with T1-weighted MRIs and manually segmented lesion masks, sourced from 33 research cohorts across 20 institutions worldwide.  In our study, we used the 655 cases with publicly accessible MRIs and GT masks. The scans were acquired using 1.5-Tesla and 3-Tesla MR scanners, with most having high resolutions (1 mm³ or higher), except for four cohorts with at least one dimension between 1–2 mm³. The lesion annotation process employed ITK-SNAP for its semi-automated tool. Annotators were trained in neuroanatomy and standardized protocols, with quality control ensured through guidance from a neuroradiologist and extensive team feedback. Each lesion was traced twice to ensure consistency. The pre-processing involved intensity standardization and linear registration of T1-weighted images and lesion masks to the MNI-152 template using the MINC toolkit.

To evaluate segmentation performance on small lesions, we utilized the MSD Lung Tumor dataset and the orCaScore dataset. The MSD Lung Tumors dataset is acquired from The Cancer Imaging Archive (TCIA), which focuses on the segmentation of NSCLC lesions from CT scans. The dataset consists of thin-section CT scans from 96 patients with NSCLC, officially divided into 64 cases for training and 32 cases for testing. In our study, we used the 64 cases with publicly available CT images and corresponding ground truth masks. The dataset was further divided into 54 patients for training and 10 patients for testing. The training set was then split into 43 patients for training and 11 patients for validation. All scans are non-contrast-enhanced chest CTs, acquired using various scanner settings. The in-plane resolution of the CT scans ranges from 0.60 mm to 0.98 mm, with slice thicknesses varying between 0.63 mm and 2.5 mm. We also included the orCaScore dataset from the MICCAI 2014 Challenge on Automatic Coronary Calcium Scoring contains CT scans from 72 patients across four European academic hospitals, acquired using scanners from different vendors. Each patient has a non-contrast enhanced, ECG-triggered cardiac calcium scoring CT (CSCT) scan. Coronary artery calcifications (CAC) were manually annotated by an experienced radiologist and a physician. Annotations were provided for calcifications with in-plane resolutions of 0.4–0.5 mm and slice thicknesses/spacings of 2.5–3 mm. The dataset is divided into 32 patients for training and 40 patients for testing (original set). However, GT masks are only available for the training set and not for the testing set. Therefore, we split the 32 patients in the training set into 26 for training, 5 for validation, and 1 for testing. In this study, after model building, the evaluation was performed using the original set of 40 patients for testing, by submitting segmentation results to the official dataset evaluation protocol ~\cite{bib34}.

\subsection{Evaluation Setup}
We used UNet++ as our segmentation model baseline due to its enhanced performance compared to U-Net in the three segmentation tasks evaluated in this study. To evaluate the performance of the GRCSF, we compared it against several state-of-the-art CNN and transformer-based methods commonly used in medical image segmentation: U-Net and UNet++, which utilize skip connection to enhance multi-scale features; Deeplabv3, which introduces various receptive fields to obtain better spatial information; TransUNet, SwinUNet, TransFuse and UTNet, which take advantage of self-attention mechanism for better global context; and SPiN which is specifically developed for stroke lesion segmentation. We also compared our method with the foundation model SAM, which benefits from pre-training on a large-scale dataset. For a fair comparison, all methods were trained and tested on the same datasets, except for SAM, which was evaluated in a zero-shot setting using bounding box prompts derived from UNet++ predictions. The comparison results on the ATLAS and MSD Lung Tumor datasets are shown in Table 1, while the results for the orCaScore dataset are presented in Table 2.

To investigate the contributions of different components in GRCSF, we conducted ablation studies using the ATLAS dataset and UNet++ as the backbone. These experiments aimed to assess the performance improvements introduced by the proposed modules of GCU, RCU and Importance Scoring mechanism. The GCU was applied at each skip connection layer in the backbone, although it can be adjusted depending on the task. Similarly, the RCU was applied to the last three upsampling layers in this study, but it can be selectively disconnected from specific layers as required. The results are summarized in Table 3. We also validated key hyperparameter choices, including the number of MAE iterations for generating residual maps, the associated inference time, the selection of MAE mask ratios, and the patch size for cross-attention in the RCU module. These experiments were conducted on one training and test subset of the ATLAS dataset.

In addition, we performed a comprehensive analysis of alternative design choices within each GRCSF module. This includes evaluating different backbone architectures, such as the lightweight MobileNet~\cite{bib69}. For the GCU, we explored variants using spatial attention modules (SAM)~\cite{bib70} instead of SE blocks, and compared placement strategies involving pyramidal attention modules (PAM)~\cite{bib71} and attention-gated (AG)~\cite{bib72} skip connections. We also assessed different strategies for residual map construction, including Grad-CAM maps generated from UNet predictions, an alternative reconstruction-based method SimMIM~\cite{bib54}, and different methods for computing the difference between the original and reconstructed images, including AbsDiff, MSE, and structural similarity index (SSIM). To demonstrate the advantage of reconstruction over simple pre-training, we additionally evaluated a model that uses a MAE-pretrained encoder to replace the original UNet encoder. The results are presented in Table 7.

For the ATLAS dataset, the four-subset design was implemented to evaluate the generalizability of the models across different medical centers and scanners while using relatively small amounts of training data compared to perform cross-validation on the entire dataset. The primary performance metric for both the ATLAS and MSD Lung Tumor dataset was the Dice Similarity Coefficient (DSC), which quantifies the similarity between the predicted and GT regions. Intersection over Union (IoU) was also reported to measure the overlap between the predicted and GT regions. Additional metrics include precision, recall, false positive rate (FPR), and volume over-segmentation error (VOSE). FPR and VOSE quantify the degree of over-segmentation relative to the background and ground-truth lesion volume, respectively. All metrics were computed on a per-patient, pixel-wise basis to ensure precise evaluation.

The orCaScore evaluation framework is a publicly accessible, web-based protocol for algorithm evaluation, originally introduced at the MICCAI 2014 workshop in Boston, USA. This framework evaluates algorithms based on the number and volume of identified calcifications on a per-patient basis. For this study, the test results were post-processed according to specific guidelines: coronary calcifications were defined as connected voxel groups (using 3D 6-connectivity) with intensities exceeding 130 Hounsfield Units (HU). To enable a more comprehensive comparison, we also report pre- and post-processed metrics using a held-out test patient separated from the training set.

\subsection{Implementation Details}
For the ATLAS dataset, the 655 cases were divided into four subsets, which is similar to the data split in~\cite{bib38}, with each subset containing patients from distinct cohorts. Within each subset, patients were further split into training and testing sets, with 80\% allocated for training and 20\% for testing. In particular, test patients were drawn from cohorts different from those in the training set. The pre-trained MAE model, enhanced with an additional GAN loss for more realistic generation (ViT-Large architecture, training mask ratios of 50\% and 75\%), was trained using all non-lesion training slices from all subsets. The training process was conducted over 200 epochs with a batch size of 32. The Adam optimizer was utilized to minimize MSE loss, with an initial learning rate of 0.001. The learning rate decayed by a factor of 0.9 every 50 epochs to ensure stable convergence. The segmentation model was trained with an input size of 224x224 using the early stopping strategy, which stops training if the validation loss did not decrease for 10 consecutive epochs. A batch size of 21 was used, and the Adam optimizer was employed with an initial learning rate of 0.0001. Optimization incorporated a cosine warmup schedule during the first five epochs, where the learning rate increased 10 times. The first and second moment estimates for the optimizer were set to 0.9 and 0.999, respectively. The weights of the convolutional filter were initialized using the He method~\cite{bib41}, and the biases were initialized to zero. For the cross-attention operation in the RCU module, the feature patch sizes at the last three decoder layers were configured as 8×8, 8×8 and 16×16, respectively. The MSD Lung Tumor dataset used the data split strategy described in the Datasets section and was trained under the same settings as the ATLAS dataset.

For the orCaScore dataset, following the data split described in the Datasets section, the same pre-trained MAE models were trained using all non-lesion training slices over 200 epochs with a batch size of 32. The Adam optimizer was employed to minimize the MSE loss, starting with an initial learning rate of 0.001, which decayed by a factor of 0.9 every 50 epochs. The segmentation network training followed the same configuration as for the ATLAS dataset, with the following adjustments: an input size of 512×512, a smaller batch size of 2 due to computational resource constraints and a reduced initial learning rate of 0.0001. Additionally, the learning rate decayed by a factor of 0.5 if the validation loss did not decrease over 5 consecutive epochs. The RCU was presented in the last three decoder layers, the feature patch sizes were configured as 16×16, 16×16 and 32×32, respectively.

All experiments were conducted on a NVIDIA RTX A6000 Ada GPU and Pytorch version 2.0.0.

\begin{table*}[ht]
\centering
\LARGE
\renewcommand{\arraystretch}{1.5} 
\setlength{\tabcolsep}{5pt} 
\resizebox{\textwidth}{!}{ 
\begin{tabular}{l|ccccccc|ccccccc|ccc}
\hline
\multirow{2}{*}{\textbf{Methods}} 
& \multicolumn{7}{c|}{\textbf{ATLAS}} 
& \multicolumn{7}{c|}{\textbf{MSD Lung Tumor}} 
& \multicolumn{3}{c}{\textbf{Model Complexity}} \\
& Dice & IoU & Precision & Recall & VOE & FPR & Inference Time & Dice & IoU & Precision & Recall & VOE & FPR & Inference Time & Model Parameters & GFLOPs & Peak Memory \\
\hline
UNet & 0.372 & 0.276 & 0.444 & 0.418 & 0.818 & 0.047\% & 2.42s & 0.670 & 0.521 & \underline{0.724} & 0.702 & 0.528 & \textbf{0.005\%} & 25.62s & 31.04M & 41.85 & 210.7MB \\
UNet++ & 0.381 & 0.286 & 0.476 & 0.409 & 0.914 & 0.054\% & 3.48s & 0.691 & \underline{0.538} & 0.643 & 0.770 & 0.471 & 0.010\% & 36.63s & 36.63M  & 106.1 & 327.8MB\\
DeepLabv3 & \underline{0.395} & 0.295 & \textbf{0.531} & 0.413 & \underline{0.485} & \underline{0.038\%} & 3.36s & 0.633 & 0.473 & 0.660 & 0.672 & 0.525 & \underline{0.006\%} & 37.28s  & 39.63M & 31.41  & 301.1MB \\
SwinUnet & 0.281 & 0.204 & 0.450 & 0.271 & 0.703 & \underline{0.038\%} & 3.86s & 0.248 & 0.148 & 0.475 & 0.198 & 0.387 & \textbf{0.005\%} & 36.84s & 41.34M & 8.69 & 212.5MB  \\
TransUnet & 0.394 & \underline{0.304} & 0.427 & 0.450 & 0.871 & 0.069\% & 4.31s & 0.680 & 0.533 & 0.657 & 0.776 & 0.689 & \underline{0.006\%} & 37.23s & 93.23M & 24.67 & 803.7MB\\
SPiN & 0.376 & 0.288 & 0.453 & 0.404 & 1.100 & 0.061\% & 3.69s & 0.683 & 0.534 & 0.621 & \textbf{0.832} & 0.766 & 0.008\% & 54.02s & 5.17M & 6.23 & 171.9MB\\
TransFuse & 0.366 & 0.274 & 0.407 & \textbf{0.476} & 1.448 & 0.077\% & 6.21s & 0.586 & 0.425 & 0.478 & \underline{0.800} & 1.071 & 0.010\% & 65.23s & 143.39M & 63.40 &737.6MB \\
UTNet & 0.344 & 0.259 & 0.466 & 0.430 & \textbf{0.465} & 0.042\% & 4.10s & 0.645 & 0.500 & 0.695 & 0.659 & \textbf{0.296} & \textbf{0.005\%} & 36.18s & 80.77M & 81.83 & 515.1MB   \\
SAM & 0.360 & 0.261 & 0.438 & 0.399 & 1.089 & \textbf{0.010\%} & 65.33s & \underline{0.722} & \textbf{0.583} & \textbf{0.745} & 0.728 & \underline{0.307} & 0.007\% & 79.50s & 631.58M & 10934.57 & 5746MB \\
GRCSF (Ours) & \textbf{0.422} & \textbf{0.319} & \underline{0.497} & \underline{0.451} & 0.942 & 0.050\% & 25.93s * & \textbf{0.730} & \textbf{0.583} & 0.709 & 0.780 & 0.370 & \textbf{0.005\%} & 61.59s * & 42.85M & 123.10 & 690.0MB \\
\hline
\end{tabular}
}
\caption{Comparison of ten segmentation methods on the ATLAS and MSD Lung Tumor test sets. The best scores are shown in bold and the second-best are underlined. ATLAS metrics are first averaged across patients within each of the four test subsets and then averaged across the subsets, whereas MSD Lung Tumor metrics are averaged over all test patients. Inference time is reported as the mean per patient. * indicates that the reported time is the end-to-end runtime of the entire framework, including the time required for MAE residual map generation and segmentation. “Model Complexity” lists the number of parameters, Giga floating-point operations per second (GFLOPs), and peak GPU memory consumption.}
\label{table:comparison}
\par
\end{table*}

\begin{table}[ht]
\centering
\large
\renewcommand{\arraystretch}{1.5}
\setlength{\tabcolsep}{4pt}
\resizebox{\textwidth}{!}{
\begin{tabular}{l|ccccc|cccc|cccc}
\hline
\multicolumn{14}{c}{\textbf{orCaScore}} \\
\hline
\multirow{2}{*}{Methods} 
& \multicolumn{5}{c|}{\textbf{Test Patients (Post-Processed)}} 
& \multicolumn{4}{c|}{\textbf{Validation Patient (Post-Processed)}} 
& \multicolumn{4}{c}{\textbf{Validation Patient (without Post-Processing)}} \\
& F1 vol & Sens vol & PPV vol & Sen lesion & PPV lesion 
& F1 vol & Sens vol & PPV vol & VOE
& F1 vol & Sens vol & PPV vol & VOE \\
\hline
UNet & 0.924 & \textbf{0.971} & 0.881 & \underline{0.871} & 0.841 & 0.889 & 0.808 & 0.989 & 0.009 & 0.743 & 0.808 & 0.687 & 0.368 \\
UNet++ & \underline{0.937} & 0.954 & 0.920 & 0.801 & 0.880 & 0.921 & 0.859 & 0.992 & 0.007 & \underline{0.843} & 0.859 & 0.828 & 0.178 \\
DeepLabv3 & 0.936 & 0.929 & \textbf{0.944} & 0.516 & \textbf{0.925} & \textbf{0.971} & \textbf{0.951} & 0.992 & 0.008 & 0.636 & \textbf{0.951} & 0.478 & 1.041 \\
SwinUnet & 0.370 & 0.251 & 0.701 & 0.275 & 0.584 & 0.467 & 0.323 & 0.843 & 0.060 & 0.400 & 0.323 & 0.525 & 0.292 \\
TransUnet & 0.770 & 0.673 & 0.902 & 0.149 & 0.840 & 0.806 & 0.730 & 0.900 & 0.081 & 0.804 & 0.730 & 0.895 & 0.085 \\
SPiN & 0.913 & 0.896 & \underline{0.930} & 0.253 & 0.892 & 0.851 & 0.746 & 0.990 & 0.009 & 0.842 & 0.746 & \textbf{0.967} & \textbf{0.033}  \\
TransFuse & 0.749 & 0.635 & 0.913 & 0.269 & \underline{0.917} & 0.811 & 0.682 & \textbf{1.000} & \textbf{0.000} & 0.513 & 0.682 & 0.410 & 0.981 \\
UTNet & 0.859 & 0.807 & 0.919 & 0.234 & 0.840 & \underline{0.939} & \underline{0.886} & \underline{0.999} & \underline{0.001} & 0.825 & \underline{0.886} & 0.771 & 0.263 \\
SAM & 0.931 & \textbf{0.971} & 0.895 & \textbf{0.898} & 0.842 & 0.913 & 0.841 & \underline{0.999} & \underline{0.001} & 0.703 & 0.841 & 0.604 & 0.552 \\
GRCSF (Ours) (Ours) & \textbf{0.946} & \underline{0.962} & \underline{0.930} & 0.827 & 0.899 & 0.885 & 0.799 & 0.992 & 0.006 & \textbf{0.856} & 0.799 & \underline{0.921}  & \underline{0.068} \\
\hline
\end{tabular}
}
\caption{Comparison of segmentation methods on the orCaScore dataset across three subsets: the 40 online-test patients after post-processing (metrics from the online evaluation framework), a held-out test patient after post-processing, and the same patient without post-processing (metrics computed from its ground-truth mask). The best results are shown in bold and the second-best are underlined.}
\label{table:orcascore}
\end{table}

\begin{table*}[ht]
\centering
\small
\renewcommand{\arraystretch}{1} 
\setlength{\tabcolsep}{5pt} 
\renewcommand{\arraystretch}{1} 
\renewcommand{\arraystretch}{1} 
\setlength{\tabcolsep}{8pt} 
\scalebox{0.8}{ 
\begin{tabular}{l|cccccc}
\hline
\textbf{Methods} & \textbf{Dice} & \textbf{IoU}  & \textbf{Precision} & \textbf{Recall} \\ 
\hline
UNet++ & 0.381 & 0.286 & 0.476 & 0.409 \\ 
GRCSF w/o RCU & 0.394 & 0.294 & 0.485 & 0.426 \\ 
GRCSF w/o Importance Score in RCU & 0.408 & 0.306 & 0.478 & 0.439 \\ 
GRCSF & \textbf{0.422} & \textbf{0.319} & \textbf{0.497} & \textbf{0.451} \\ 
\hline
\end{tabular}
}
\caption{Ablation study results on the 4 test sets of the ATLAS dataset. The best results are highlighted in bold. Metrics are obtained by averaging the results across the 4 test sets.}
\label{table:ablation}
\par
\end{table*}

\subsection{Results}
\subsubsection{State-of-the-Art Comparisons}
For the ATLAS dataset, our GRCSF achieved a DSC of 0.422, outperforming the previous state-of-the-art method DeepLabv3, which ranked second, by 2.7\% (Table 1). Furthermore, our method demonstrated a dominant performance in IoU of 0.319, representing improvements of 1.5\% over the second highest metrics obtained by TransUnet. GRCSF achieved the second-highest precision and recall scores, suggesting balanced segmentation performance that neither oversegments nor undersegments compared to the other methods. 

For the MSD Lung Tumor dataset, our method achieved a DSC of 0.730, outperforming the foundation model SAM by 0.08\% and surpassing UNet++, the third-best performer by 3.9\% (Table 1). In terms of IoU, our method reached a score of 0.583, matching that of SAM and exceeding the second-highest UNet++ by 4.5\%. Furthermore, GRCSF obtained the lowest FPR, indicating a reduced tendency to over-segment background regions.

In terms of model complexity, the total prediction time per patient for the entire GRCSF framework is 25.93 seconds on the ATLAS dataset and 61.59 seconds on the MSD Lung Tumor dataset. This end-to-end runtime consists of two components: the segmentation network, which requires 4.07 seconds for ATLAS and 34.93 seconds for MSD Lung Tumor; and the residual map generation process, which adds 21.86 seconds (Table 4) and 26.66 seconds, respectively, when using five MAE iterations and two mask ratios. GRCSF has 42.85M parameters, similar to SwinUNet with 41.34M and slightly higher than UNet++ with 36.63M, while remaining lighter than Transformer-based models such as TransUnet with 93.23M and TransFuse with 143.39M. Its GFLOPs reach 123.10, which is higher than those of CNN-based models like UNet (41.85) and DeepLabv3 (31.41), but much lower than that of SAM, which has 10,934.57 GFLOPs. GRCSF’s peak memory usage is 690.0 MB, lower than that of TransUnet (803.7 MB) and TransFuse (737.6 MB). Overall, our method strikes a favorable balance between efficiency and accuracy.

\begin{figure*}[htbp]
    \centerline{\includegraphics[width=1.02\textwidth]{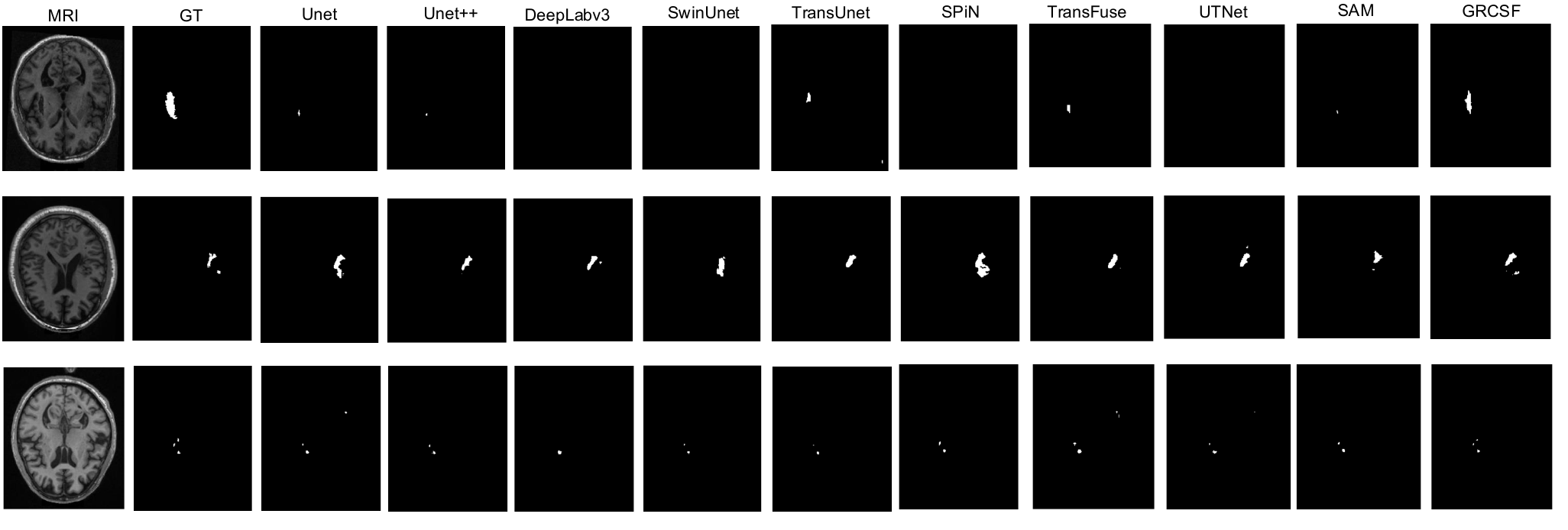}}
    \caption{Visual comparison of our proposed method GRCSF against previous state-of-the-art at medical image segmentation using ATLAS dataset. Please zoom in for a clearer view.}
    \label{Fig. 6}
\end{figure*}

\begin{figure*}[!t]
    \centerline{\includegraphics[width=1.02\textwidth] 
    {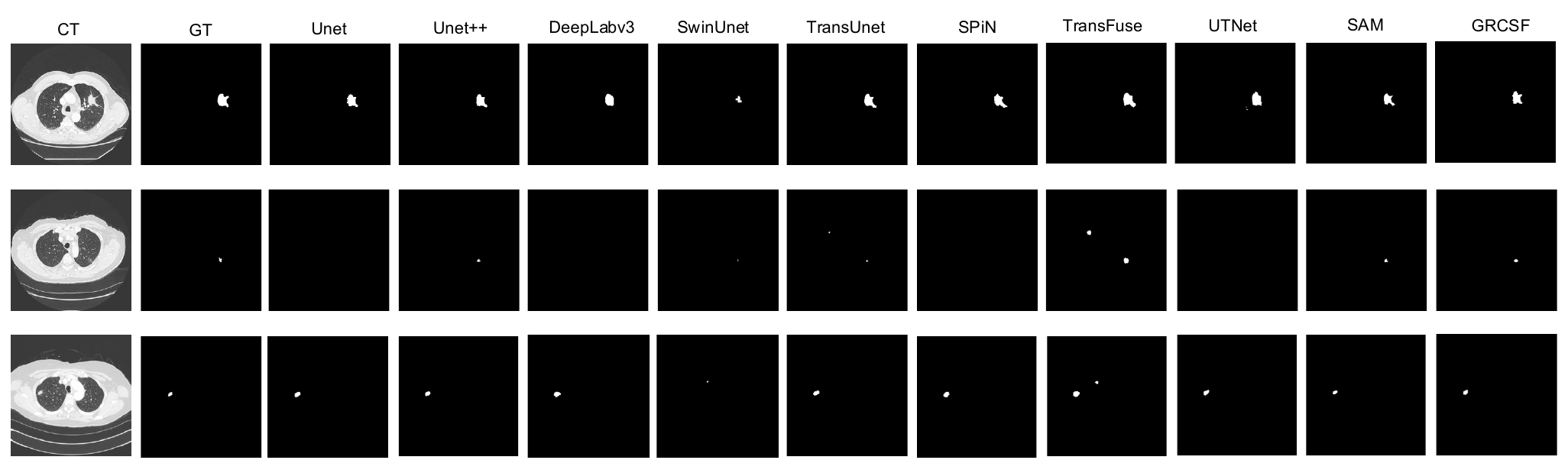}}
    \caption{Visual comparison of our proposed method GRCSF, against previous state-of-the-art methods in medical image segmentation using the MSD Lung Tumor dataset. Please zoom in for a clearer view.}
    \label{Fig. lung}
\end{figure*}

\begin{figure*}[!t]
    \centerline{\includegraphics[width=1.02\textwidth] 
    {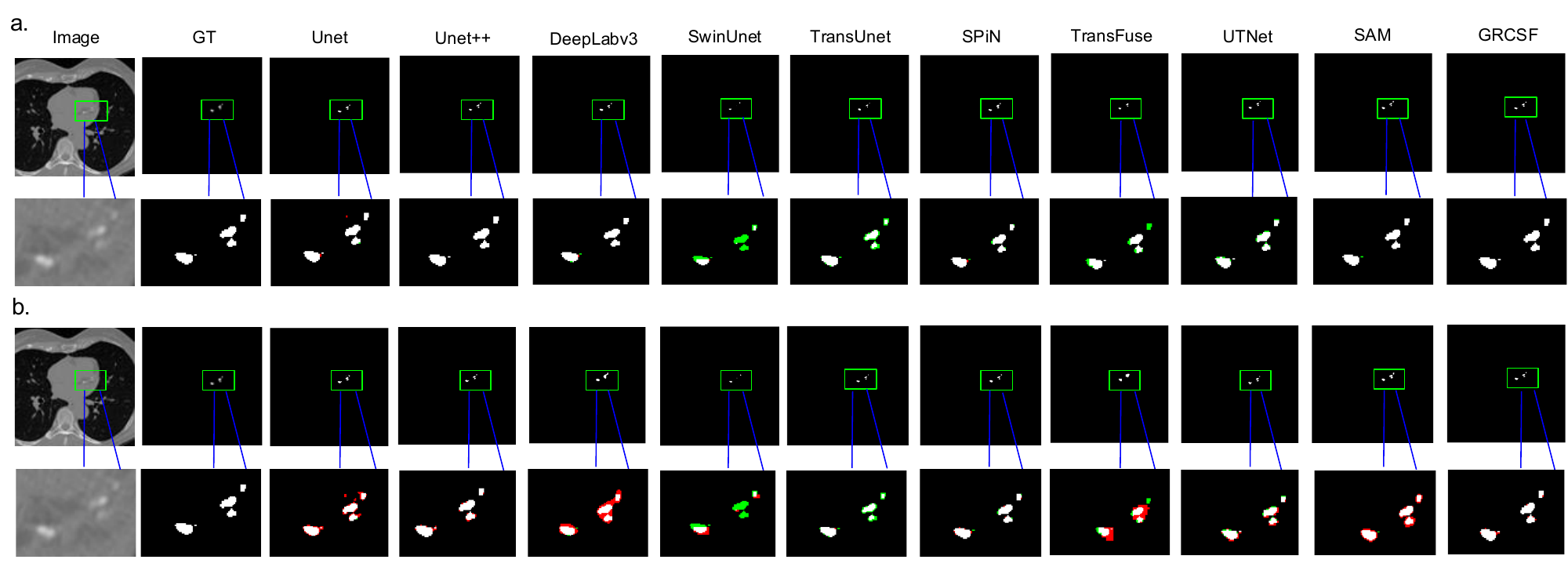}}
    \caption{Visual comparison of our proposed method GRCSF, against previous state-of-the-art methods in medical image segmentation using the orCaScore dataset. (a) Visualization of predictions after post-processing, leveraging domain knowledge that calcifications are identified as regions where the predicted image pixel values exceed 130 HU. (b) Visualization of raw predictions without post-processing. Error maps are included to highlight prediction errors on small lesions more clearly: red indicates false-positive pixels, while green indicates false negatives. Please zoom in for a clearer view.}
    \label{Fig. 7}
\end{figure*}

\begin{figure*}[htbp]
    \centerline{\includegraphics[width=0.45\textwidth]{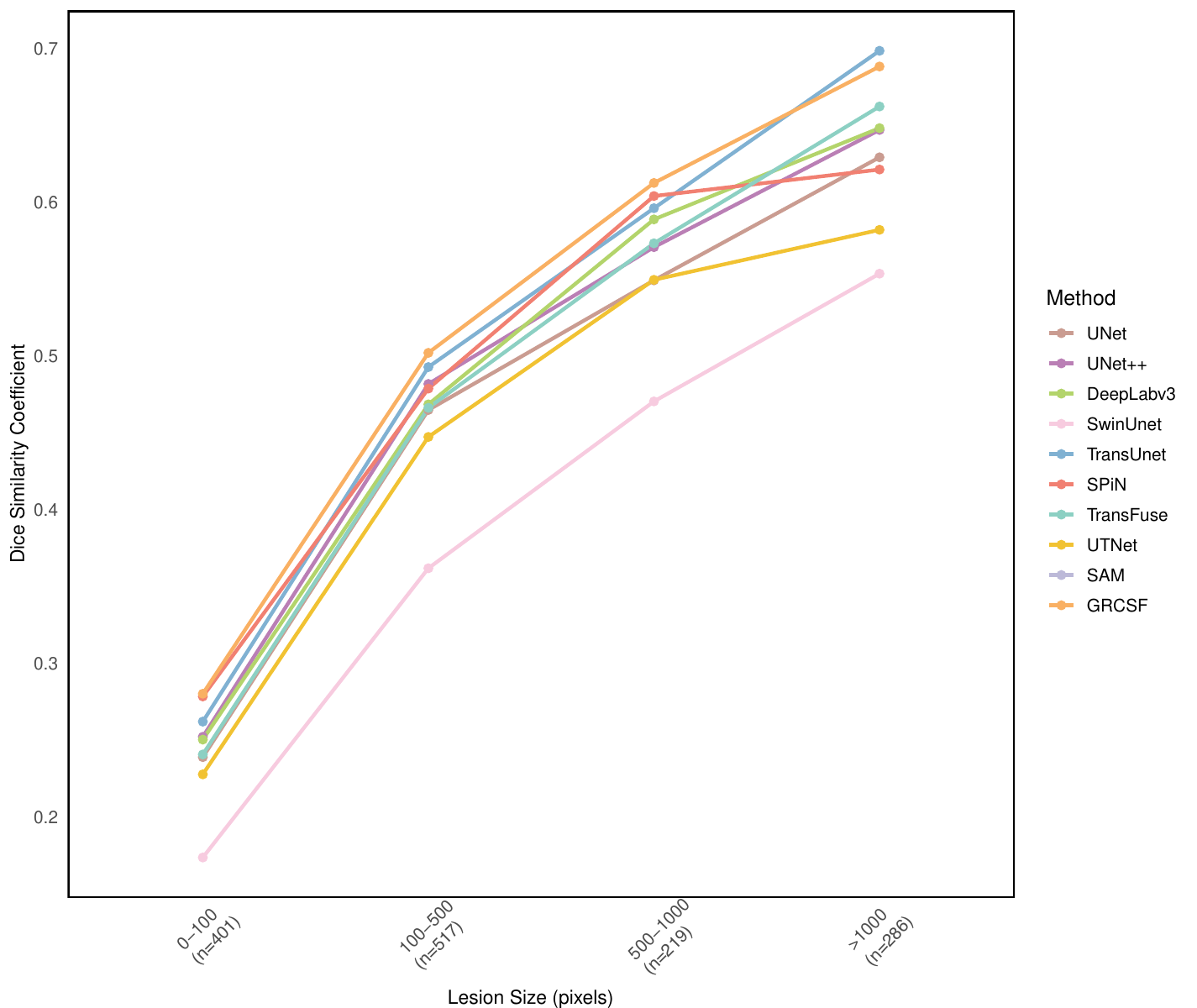}}
    \caption{Case-wise lesion-size performance of all methods on one ATLAS test subset. Please zoom in for a clearer view.}
    \label{Fig. lesion_size}
\end{figure*}

Selected example segmentation results from ATLAS, MSD Lung Tumor and orCaScore are shown in Fig. \ref{Fig. 6}, Fig. \ref{Fig. lung} and Fig. \ref{Fig. 7}. The visualizations demonstrate the effectiveness of GCU and RCU, as most methods tended to under-segment for challenging cases, whereas our method successfully captured the structures that other methods missed. As shown in Fig. \ref{Fig. lesion_size}, the case-wise performance breakdown by lesion size demonstrates that our method outperforms all others on small and ambiguous lesions ranging from 1 to 1,000 pixels. TransUnet achieves the best performance in the subgroup of lesions larger than 1,000 pixels, however our method ranks the second in that category and when averaged across all the sizes, our method is the best. The orCaScore results include both post-processed and non-post-processed visualizations of the same prediction with error maps, providing additional insights into model performance. In Fig. \ref{Fig. 7}a, the post-processed results of all methods demonstrate accurate lesion boundaries in certain methods; however, this improvement is partly due to the removal of calcifications with image pixel values below 130HU during post-processing. This artificially improves the appearance of over-segmentation methods like U-Net and DeepLabv3, masking their inherent limitations. In contrast, the non-post-processed results in Fig. \ref{Fig. 7}b reveal GRCSF’s true advantage of accurately segmenting small calcifications without significant over-segmentation. This highlights our model’s ability to capture challenging structures while maintaining recall without relying on post-processing. 

\begin{figure*}[!t]
    \centerline{\includegraphics[width=0.5\textwidth]{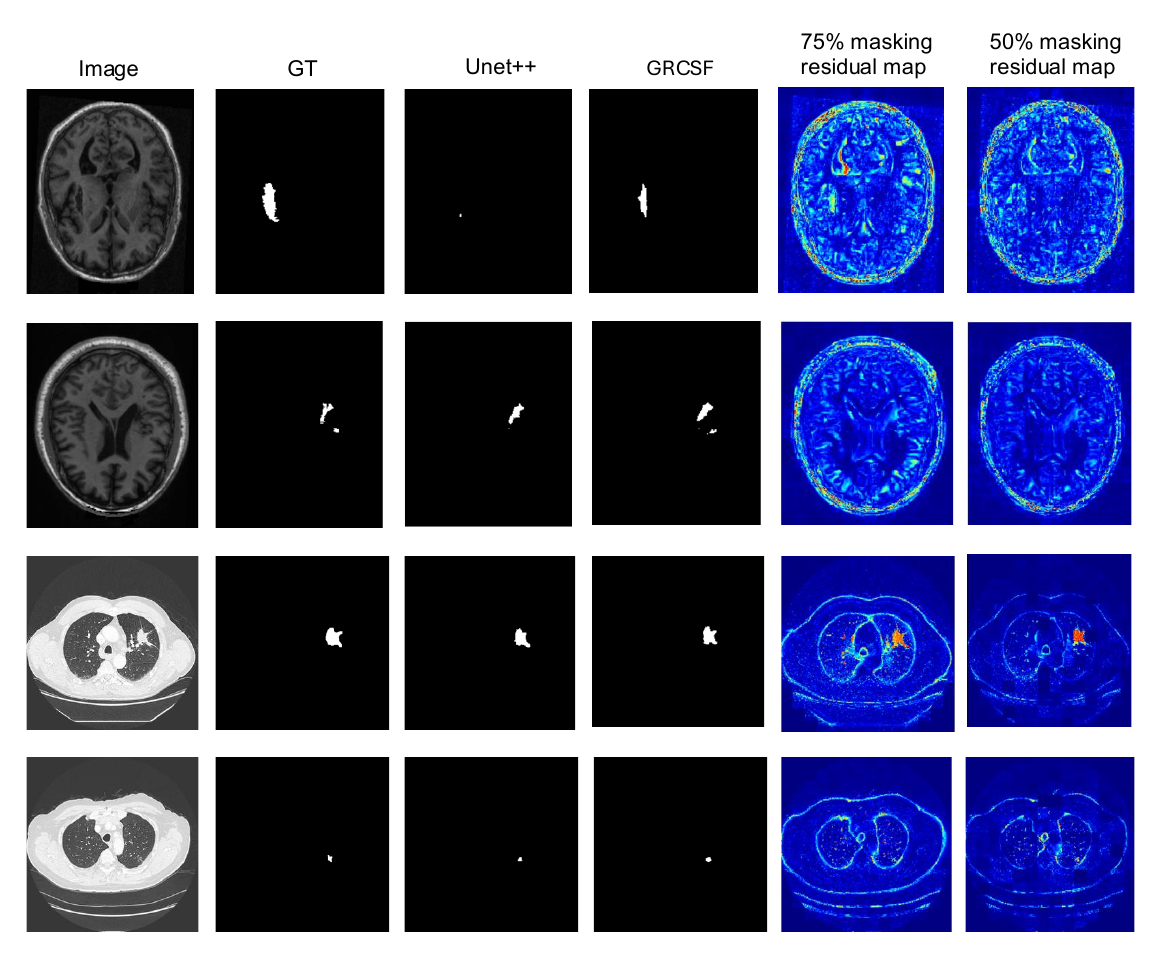}}
    \caption{Example outputs from UNet++ and our proposed method GRCSF for the ATLAS and the MSD Lung Tumor datasets. The figure includes GT segmentation and the corresponding MRM of the input images for comparison. Please zoom in for a clearer view.}
    \label{Fig. MRM}
\end{figure*}

The impact of integrating MRM on segmentation performance is illustrated in Fig. \ref{Fig. MRM}. The segmentation of the challenging lesion improved consistently when MRM was incorporated, particularly when using UNet++ as the backbone. This configuration showed the best performance across three datasets.

Segmentation outputs from UNet++ revealed common errors, such as completely missing small lesions and under-segmentation (false negatives) for lesions with subtle appearance. Low imaging contrast or insufficient global context often led to these inaccuracies. In contrast, the proposed method with MRM guidance avoided these problems and produced more accurate segmentations. 

\begin{table}[ht]
\centering
\scriptsize
\renewcommand{\arraystretch}{1} 
\setlength{\tabcolsep}{5pt} 
\renewcommand{\arraystretch}{1} 
\begin{tabular}{c|c c c c|c}
\toprule
\textbf{MAE Iterations} & \textbf{Dice} & \textbf{IoU} & \textbf{Precision} & \textbf{Recall} & \textbf{Inference Time} \\
\midrule
1 & 0.520 & 0.400 & 0.561 & 0.568 & 4.33s \\
2 & 0.554 & 0.421 & 0.557 & \textbf{0.656} & 8.71s \\
3 & 0.534 & 0.404 & 0.505 & 0.638 & 12.30s \\
4 & 0.550 & 0.417 & 0.582 & 0.646 & 18.24s \\
5 & \textbf{0.581} & \textbf{0.448} & 0.595 & 0.645 & 21.86s \\
6 & 0.573 & 0.439 & \textbf{0.603} & 0.636 & 24.08s \\
\bottomrule
\end{tabular}
\caption{Performance of GRCSF with residual maps reconstructed from different numbers of MAE iterations, evaluated on one ATLAS test subset. Inference time is reported as the mean per patient while generating residual maps for two MAE mask ratios. The best results are highlighted in bold.}
\label{table:mae iterations}
\end{table}

\begin{table}[ht]
\centering
\scriptsize
\renewcommand{\arraystretch}{1} 
\setlength{\tabcolsep}{5pt} 
\begin{tabular}{c|c c c c|c c c c}
\toprule
\textbf{Experiment} & 30\% & 50\% & 75\% & 90\% & Dice & IoU & Precision & Recall \\
\midrule
1 & \checkmark & & & & 0.537 & 0.413 & 0.557 & 0.627 \\
2 & & \checkmark & & & 0.574 & 0.439 & 0.569 & \textbf{0.681} \\
3 & & & \checkmark & & 0.541 & 0.408 & \textbf{0.601} & 0.636 \\
4 & & & & \checkmark & 0.531 & 0.403 & 0.580 & 0.572 \\
5 & & \checkmark & \checkmark & & \textbf{0.581} & \textbf{0.448} & 0.595 & 0.645 \\
\bottomrule
\end{tabular}
\caption{Performance of GRCSF using residual maps generated with different MAE mask ratios, evaluated on one ATLAS test subset. The best results are highlighted in bold.}
\label{table:mae ratios}
\end{table}

\begin{table}[ht]
\centering
\scriptsize
\renewcommand{\arraystretch}{1} 
\setlength{\tabcolsep}{5pt} 
\renewcommand{\arraystretch}{1} 
\begin{tabular}{l|c c c c c c}
\toprule
\textbf{Cross Attention Patch Size} & \textbf{Dice} & \textbf{IoU} & \textbf{Precision} & \textbf{Recall} & \textbf{Model Parameters} \\
\midrule
4×4, 8×8, 16×16 & 0.500 & 0.373 & 0.426 & \textbf{0.734} & 42.75M \\
8×8, 4×4, 16×16 & 0.482 & 0.357 & 0.408 & 0.727 & 42.82M \\
8×8, 8×8, 8×8 & 0.480 & 0.358 & \textbf{0.604} & 0.492 & 42.75M \\
8×8, 16×16, 16×16 & 0.553 & 0.415 & 0.585 & 0.583 & \textbf{42.95M}  \\
8×8, 8×8, 16×16 (Ours) & \textbf{0.581} & \textbf{0.448} & 0.595 & 0.645 & 42.85M\\
\bottomrule
\end{tabular}
\caption{Performance of GRCSF with different patch sizes for cross-attention in the RCU, evaluated on one ATLAS test subset. The best results are highlighted in bold.}
\label{table:patch size}
\end{table}

\subsubsection{Ablation Study}
Adding the GCU to skip connections improved DSC and Recall by 1.3\% and 1.7\%, respectively from the UNet++ baseline (Table \ref{table:ablation}). The RCU introduced two key operations: (1) applying cross-attention between decoder feature maps and the MRM, and (2) generating patch-based importance scores. Building on the GCU results, adding cross-attention with MRM further improved DSC and Recall by another 1.4\% and 1.3\%. Incorporating patch-based importance scores provided an additional boost of 1.4\% and 1.2\%.

Overall, GRCSF improved DSC, IoU, Precision, and Recall by 4.1\%, 3.3\%, 2.1\%, and 4.2\%, from the baseline, respectively. These results demonstrate that the integration of all proposed modules significantly enhances the framework’s segmentation performance, validating the design choices of GRCSF.

\begin{table}[ht]
\centering
\small
\renewcommand{\arraystretch}{1} 
\setlength{\tabcolsep}{5pt} 
\resizebox{\textwidth}{!}{
\begin{tabular}{ll|ccccc}
\toprule
\textbf{Module} & \textbf{Method} & \textbf{Dice} & \textbf{IoU} & \textbf{Precision} & \textbf{Recall} & \textbf{Model Parameters} \\
\midrule

\multirow{4}{*}{Residual Map in RCU} 
  & Grad-CAM & 0.476 & 0.359 & 0.470 & 0.566 &  42.85 M \\
  & SimMIM Residual Map (AbsDiff) & 0.553 & 0.421 & 0.537 & \textbf{0.663}  & 42.85 M \\
  & MAE Residual Map (MSE) & 0.547 & 0.422 & 0.541 & 0.648 & 42.85 M \\
  & MAE Residual Map (SSIM) & 0.529 &	0.404 &	0.526 &	0.633 & 42.85 M \\
  & MAE Residual Map (AbsDiff) (Ours) & \textbf{0.581} & \textbf{0.448} & \textbf{0.595} & 0.645 & 42.85 M \\

\midrule
\multirow{4}{*}{GCU} 
  & UNet++ with PAM & 0.490 & 0.399 & 0.558 & 0.613 & 72.21M \\
  & UNet++ with AG skip connections & 0.521 & 0.413 & 0.513 & 0.618 & 40.12M \\
  & SAM in GCU & 0.526 & \textbf{0.415} & 0.548 & \textbf{0.635} & 42.20M \\
  & GCU (Ours) & \textbf{0.534} & 0.408 & \textbf{0.560} & 0.615 & 42.27M \\

\midrule
\multirow{3}{*}{Segmentation Backbone} 
  & UNet with pre-trained MAE Encoder & 0.517 & 0.395 & 0.559	& 0.583 & 304.93M \\
  & GRCSF with MobileNet Encoder & 0.472 & 0.358 & 0.490 & 0.549 & 35.13M \\
  & GRCSF (Ours)  & \textbf{0.581} & \textbf{0.448} & \textbf{0.595} & \textbf{0.645} & 42.85 M \\

\bottomrule
\end{tabular}
}
\caption{Impact of different module design choices on segmentation performance, evaluated on one ATLAS test subset. The best results within each module group are highlighted in bold.}
\label{table:design choices}
\end{table}

\subsubsection{Design Choices of GRCSF Modules}
The results in Table \ref{table:mae iterations} indicate that our framework achieves the best performance when residual maps are generated using five MAE iterations. Although this configuration introduces an additional 17.53 seconds compared to using a single iteration, it yields improvements of 6.1\% in Dice and 4.8\% in IoU. Table \ref{table:mae ratios} shows that, when using a single mask ratio, 50\% and 75\% are the most effective. Combining these two ratios leads to the best overall performance in terms of Dice and IoU. As shown in Table \ref{table:patch size}, the cross-attention configuration with patch sizes of 8×8, 8×8, and 16×16 outperforms other combinations.

\begin{figure*}[htbp]
    \centerline{\includegraphics[width=\textwidth]{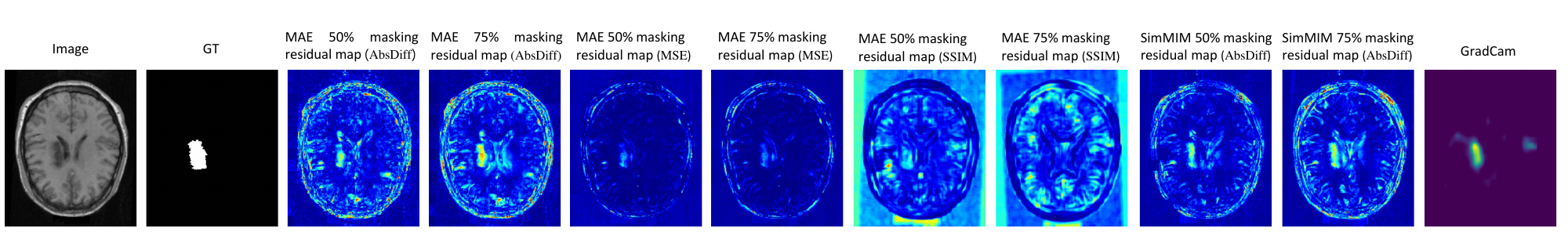}}
    \caption{Visual comparison of residual maps produced by different methods: MAE residuals computed with absolute difference (AbsDiff), MSE, and structural similarity (SSIM); SimMIM residuals at 50\% and 75\% mask ratios using AbsDiff; and the Grad-CAM map from UNet. Please zoom in for a clearer view.}
    \label{Fig. choice_of_design}
\end{figure*}

The residual maps generated by SimMIM and MAE using AbsDiff, MSE, SSIM, and Grad-CAM are visualized in Fig. \ref{Fig. choice_of_design}. Among them, the MAE residual map obtained with AbsDiff achieves the highest Dice score, followed by SimMIM. Replacing the GCU with PAM and AG skip connections does not improve performance. Similarly, substituting the SE module in our GCU design with SAM does not yield better results. Moreover, replacing the encoder of UNet++ with a lightweight architecture such as MobileNet leads to a significant drop in performance. Additionally, implementing MAE pre-trained encoder for segmentation does not demonstrate notable improvements. The results are shown in Table \ref{table:design choices}.

\section{Discussion}
Our results indicate that GRCSF consistently delivers superior segmentation performance across diverse datasets by leveraging a dual-feature compensation strategy. It integrates global feature recovery through the GCU to address downsampling losses and enhances regional features with the RCU using SSL residual maps and a patch-based cross-attention mechanism. This design enables GRCSF to effectively handle general lesion segmentation, including challenging tasks such as low-contrast and small lesion segmentation.

CNN-based models like U-Net and UNet++ rely on fixed skip connections that cannot adaptively refine feature maps, resulting in the loss of detailed features. Their fixed receptive fields further limit their ability to capture global contextual information required for segmenting lesions of varying scales and locations. Consequently, U-Net tends to over-segment lesions or miss small lesions altogether. While UNet++ mitigates over-segmentation compared to U-Net, it still struggles to detect challenging lesions with sufficient sensitivity. Similarly, SPiN improves resolution by upsampling input images but struggles with accurate segmentation due to its tendency to over-segment, leading to false positives. SPiN lacks mechanisms to dynamically focus on critical regions, which further limits its precision in complex cases. GRCSF effectively addresses the limitations of the above-mentioned methods in challenging lesion segmentation tasks. It significantly improves sensitivity to small and low-contrast lesions by feature compensation. GRCSF also resolves the blurred edges and ambiguous segmentation common in the output produced by other methods. Additionally, by dynamically focusing on lesion-specific regions, GRCSF minimizes false positives, enhancing the reliability of segmentation results in complex cases. Transformer-based models such as TransUNet, TransFuse, SwinUNet and UTNet attempt to capture long-range dependencies and enhance global spatial information. However, their patch-based designs often struggle to capture detailed context around small lesions and fail to provide sufficient edge sensitivity, resulting in high false-negative rates. Additionally, due to their high model complexity, these methods are prone to overfitting and yield suboptimal results on small datasets such as orCaScore, thereby limiting their practicality and generalizability. In contrast, GRCSF integrates global and regional features effectively without requiring complex parameter tuning. It achieves higher recall in lesion segmentation while avoiding over-segmentation. The model’s superior performance, even without post-processing of orCaScore results, further highlights its practicality for real-world applications where post-processing conditions are often limited. Notably, SAM does not outperform our method on any of the three datasets. However, it demonstrates effectiveness in refining high-contrast and irregular lesions, such as lung tumors, as evidenced by its improved performance over UNet++ on the MSD Lung Tumor dataset when refining the bounding boxes derived from UNet++ predictions. In contrast, it struggles with low-contrast brain lesions, where the same prompts result in worse performance than UNet++. Moreover, SAM lacks the ability to correct false positives at the lesion level; as a result, when prompts frequently include such cases, it leads to lower precision and higher FPR and VOSE. Its performance could potentially be improved through few-shot learning and advanced prompt engineering strategies.

In particular, the ATLAS dataset posed unique challenges due to significant image variability, including differences in imaging protocols, scanner types, and cohort demographics across medical centers. In our study, certain subsets involved training on a few cohorts while testing on entirely different cohorts, reflecting differences in cohort characteristics. Despite these challenges, GRCSF demonstrated exceptional robustness, consistently achieving higher segmentation performance across all four subsets, thereby validating its strong generalizability under diverse training and testing conditions. While our method incorporates residual maps generated by MAE, which introduces additional computational cost, this increase remains within practical limits. Specifically, GRCSF requires only 23.51 and 35.97 seconds more than the fastest baseline UNet per patient on the ATLAS and MSD Lung Tumor datasets, respectively, yet achieves Dice improvements of 5\% and 6\%. In time-sensitive scenarios, the full pipeline can process a patient in approximately 1 minute on a single A6000 GPU, offering a favorable trade-off between accuracy and efficiency for routine deployment.

The observed improvements in the ablation study highlight the complementary roles of global and regional compensation in GRCSF. The effective utilization of pixel-level feature similarity from different sources helps the model focus on regions that are often overlooked. The 50\% and 75\% mask ratios in MAE generate complementary MRM, which are adaptively integrated through learnable weights and cross-attention mechanism. This integration dynamically highlights informative regions while suppressing irrelevant ones, thereby mitigating false negatives and false positives caused by reconstruction errors. By leveraging these MRM through the RCU, GRCSF prioritizes regions with high lesion likelihood and compensates for missed feature representations, which is particularly effective for low-contrast and small lesions. This demonstrates the added value of MRM in comparison to baseline networks, which exhibited suboptimal performance in these challenging scenarios.

Unlike traditional SSL methods that use MAE for pre-training, our method directly utilizes the reconstructed images to generate SSL residual maps. While involving domain and high-level image features, preserving more low-level and pixel-level information. This method aligns with the learnable and consistent anatomical structures across patients. These benefits are not achievable through pre-trained encoders alone. Although the MAE pre-training configuration outperforms the original UNet baseline, it still falls short of the performance achieved by the full GRCSF pipeline. Moreover, using a fixed MAE encoder limits architectural flexibility. In contrast, components such as the GCU can be integrated into U-shaped backbones but become more complex to implement with a frozen MAE encoder. Moreover, averaging five runs of MRM generation mitigates the variability introduced by random masking, ensuring reliable guidance for the backbone. This strategy significantly improves segmentation accuracy and robustness, offering a novel way to integrate SSL into medical image analysis. An alternative of MAE MRM is SimMIM MRM, which appears to produce smoother reconstructions and visually sharper structural details, which makes the resulting residual map less suitable for lesion segmentation, especially for small lesions. This is because the SimMIM reconstruction allows normal anatomical structures to dominate the residual, obscuring the regions that contain small lesions. In contrast, MAE captures global morphological features throughout the dataset and leaves larger reconstruction errors over abnormal tissues, which more effectively highlight lesion areas.

\section{Conclusion}
In this work, we propose GRCSF, a dual-feature compensation framework that leverages a U-shaped backbone and SSL residual maps to improve lesion segmentation. We evaluate its effectiveness using three example datasets representing low-contrast and small lesion segmentation tasks. The GCU mitigates information loss caused by downsampling in the encoder, enhancing global feature representation to improve boundary delineation and small lesion detection. The RCU integrates complementary pixel-level features from MRM, employing a patch-based importance scoring mechanism to localize lesions more effectively while reducing false positives. Together, GRCSF provides a robust and efficient solution, demonstrating improvement in challenging medical imaging segmentation tasks.

{\bfseries Limitation:} 
The overall computational cost, including residual map generation and the use of a moderately sized backbone, may present challenges for large-scale deployment or routine clinical use. In our future work, we will aim to reduce model complexity by exploring efficient strategies for residual map generation, for example, replacing random masking with a more targeted approach that applies masks to regions of interest. In addition, the current UNet++ backbone could be substituted with a lightweight yet high-performance alternative to further reduce training and inference cost.

\section{Acknowledgments}
This work was supported by the Australian Research Council Discovery project (DP200103748), NHMRC Investigator (APP2017023) and in part by Startup funds from The University of Newcastle.





\end{document}